# Luminosity Measurement at ILC

I. Bozovic-Jelisavcic[1*] and H. Abramowicz[2], P. Bambade[3], T. Jovin[1], M. Pandurovic[1], B. Pawlik[4], C. Rimbault[3], I. Sadeh,[2] I. Smiljanic[1]

1 – Vinca Institute of Nuclear Sciences, University of Belgrade
Belgrade – Serbia

2 – School of Physics and Astronomy, Tel Aviv University
Tel Aviv – Israel

3 – LAL, University Paris-Sud, IN2P3/CNRS
Orsay – France

4 – Institute of Nuclear Physics PAN
Cracow – Poland

More than twenty institutes join the FCAL Collaboration in study of design of the very forward region of a detector for ILC and CLIC. Of particular importance is an accurate luminosity measurement to the level of $10^{-3}$, a requirement driven by the potential for precision physics at a future linear collider. In this paper, the method for luminosity measurement, requirements on luminometer and its integration in the forward region are presented. The impact of several effects contributing to the systematic uncertainty of luminosity measurement is given.

## 1 Introduction

Physics requirements like production cross-sections measurements, anomalous gauge boson couplings, EW physics and new physics searches impose the high precision of luminosity measurement at a future linear collider. Luminosity will be measured from Bhabha scattering that is dominantly QED process at ILC energies. Achievable precision is limited both by the reconstruction of scattered Bhbha particles as well as by physics effects which have to be experimentally controlled (beam-beam effects, presence of physics background). At the same time, there is an on-going theoretical effort to complete NLO corrections of the Bhabha cross-section.

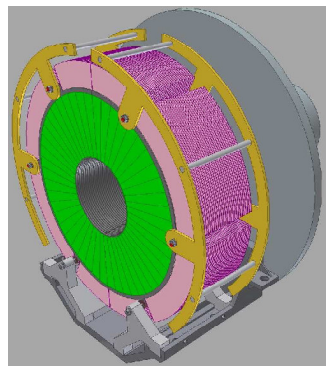

Figure 1: Mechanical structure of the luminosity calorimeter.

## 2 Luminometer at ILC

Luminosity calorimeter is foreseen as a sampling silicon/tungsten calorimeter consisting of 30 one radiation length thick absorber planes followed by segmented silicon sensor planes. To keep the Moliere radius of 1.5 cm, 1 mm sensor gaps are provided. As illustrated in Figure 1, tungsten disks are precisely positioned using 4 bolts. The system is additionally stabilized by steel rings. Reconstruction of the polar angle of electron is influenced by sensor segmentation that is optimized to 48/64 azimuthal/radial divisions. Luminosity calorimeter is positioned 2.5 m from the IP, with the geometrical aperture between 31 mrad and 78 mrad. Sufficient statistics corresponding to the 2.1 nb of integrated cross-section of the signal is obtained in the detector fiducial volume defined within [41,69] mrad [1]. By restricting the signal for luminosity measurement to the detector fiducial volume, only events with no shower leakage through the edges of luminosity calorimeter are selected. In this volume, a stable sampling term $\alpha_{res}$ (2.1), usually referred to as 'energy resolution', is obtained (Figure 2).

---

[*] This study has been supported by: Ministry of Science and Technology of the Republic of Serbia through the Project No. 15004B and also by the Commission of the European Communities under the 6[th] Framework Program "Structuring the European Research Area", contract number RII3-026126.

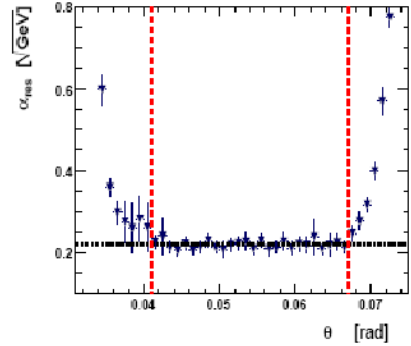

Figure 2: Energy resolution for 250 GeV electrons as a function of the polar angle. Dashed lines mark fiducial volume of the luminosity calorimeter.

In (1.1), the usual parameterization of energy resolution $\sigma_E$ corresponding to the standard deviation of an energy distribution with a mean E, deposited by electron beam $E_{beam}$ is given:

$$\frac{\sigma_E}{E} = \frac{\alpha_{res}}{\sqrt{E_{beam}[GeV]}} \qquad (2.1)$$

Simulating only electron showers inside the luminometer's fiducial volume, $\alpha_{res}$ is estimated to be: $\alpha_{res} = (0.21 \pm 0.02_{stat})\sqrt{GeV}$. Parameter $\alpha_{res}$ is found to be independent of the shower energy in the range from 50 GeV to 300 GeV. In the same range, the response of luminosity calorimeter is linear with respect to the shower energy [1]. The position of an EM shower in the detector is reconstructed by performing a weighted average over the depositions on individual pads. The weight $W_i$, of a given detector pad $i$, is determined by logarithmic weighting [2], for which $W_i = max\{0, C + \ln(E_i/E_{tot})\}$. The symbol $E_i$ refers to the individual pad energy, $E_{tot}$ is the total energy in all pads, and C is a constant. In this way, only pads which contain a sufficient fraction of the shower energy contribute to the reconstruction. The polar angle resolution $\sigma_\theta$, and a polar angle measurement bias $\Delta\theta$, are defined as the Gaussian width and the central value of the distribution of the difference between reconstructed and generated polar angles. They are found to be $(2.2 \pm 0.01) \cdot 10^{-2}$ mrad and $(3.2 \pm 0.1) \cdot 10^{-3}$ mrad, respectively. Uncertainties of $\alpha_{res}$, $\sigma_\theta$ and $\Delta\theta$ will be considered in Chapters 3.2.3 and 3.2.4 as sources of systematic uncertainty for the luminosity measurement.

# 3  Luminosity measurement an ILC

## 3.1  Method

At ILC, the integrated luminosity will be determined from the counted number of Bhbaha events reconstructed in the detector fiducial volume $N_{exp}$, corrected for the number of miss-counted events due to various systematic effects. As shown in (3.1), the measured luminosity will also depend on the selection efficiency $\varepsilon$ and the theoretical cross-section for Bhabha scattering $\sigma_B$.

$$L_{int} = \frac{N_{exp} - \sum_i N_i^{cor}}{\varepsilon \cdot \sigma_B}$$ (3.1)

In order to exploit the characteristic topology of Bhabha events with two back-to-back showers deposited almost full beam energy in forward and backward arms of the detector and, at the same time, minimize the suppression of the Bhbaha cross-section due to beam-beam effects, the following empirical selection is applied [3]: the polar angle of the reconstructed shower must be within the detector fiducial volume [$\theta_{min}$, $\theta_{max}$] at one side and within [$\theta_{min}$+4 mrad, $\theta_{max}$-7mrad] at the other, and the total energy deposited in the luminometer must be more than 80% of the center-of-mass energy. Polar angle criterion is subsequently applied at the forward and backward side of the detector in order to avoid systematic bias from the longitudinal position of the interaction point.

## 3.2 Systematic uncertainties

### 3.2.1 Beam-beam effects

The acceleration of electrons and positrons towards the bunch center when bunches are crossing changes their momentum and, more importantly, electrons and positrons radiate beamstrahlung prior to Bhabha scattering. In addition, final state Bhabha particles get focused by the electromagnetic field of the opposite space charge. The result is an effective reduction of the Bhabha cross-section in the detector fiducial volume, with the dominant contribution stemming from the beamstrahlung. Size of this Bhabha Supresion Effect (BHSE) is found to depend on selection criteria for the luminosity measurement, amounting to 1.51%±(0.05%)$_{stat}$ [3]. for nominal beam parameters at 500 GeV center of mass energy, for the proposed event selection. BHSE can be understood as an effect one can correct for, once its experimental uncertainty is known. Data-driven method to measure the beamstrahlung component of BHSE has been proposed, based on the reconstruction of the luminosity spectrum [3]. Experimental uncertainty is resulting from the precision to which bunch sizes $\sigma_x$ and $\sigma_z$ are measured. In this paper, a BHSE experimental uncertainty is chosen to correspond to the 5% relative error of the bunch size measurement.

### 3.2.2 Background from physics processes

Four-fermion production via Neutral Current is known to have a large cross-section with maxima at low polar angles. It is dominated by the multiperipheral Feynman diagram where two virtual photons are exchanged between electron spectators. The spectators remain at high energy. Less than 1% of spectators hits the luminosity calorimeter and manifests as a background for Bhabha events. The cross-section amounts to (12.0±0.5) nb at 500 GeV assuming photons with momentum larger than 0.1 GeV/c being exchanged. We used the WHIZARD [4] event generator to obtain sample of events for final states with leptons in the inner legs. Event generator is tuned to reproduce LEP data for charm production in two-photon processes [5] by adjusting the minimal exchanged momentum of the photon to $10^{-4}$ GeV/c. A Bhabha sample of 5 pb$^{-1}$ has been generated with the cross-section of (4.70±0.03) nb, at 500 GeV, using the BHLUMI [6] event generator. As mentioned in 3.1, the event selection is optimized to reduce Bhabha suppression from space charge effects. After the selection is applied, the overall impact of four-fermion events on luminosity measurement saturates around background to signal ratio B/S=2.3 $10^{-3}$ at 500 GeV. The Bhabha event selection efficiency is sufficient to maintain statistical error below $10^{-3}$ for annual running at 500 GeV and at nominal luminosity. Projection of background hits on the front plane of the luminosity calorimeter is shown in Figure 3, before and after event selection is applied.

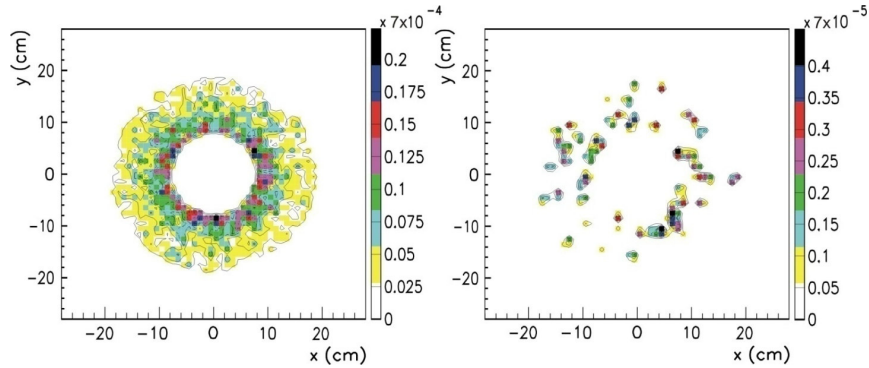

Figure 3: Projections of background hits in the luminometer front plane before *(left)* and after the event selection *(right)*.

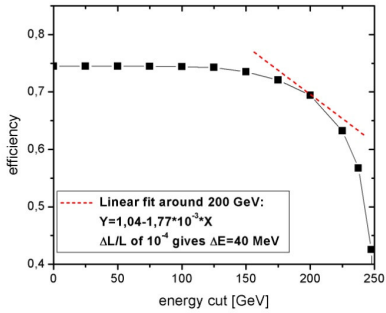

Figure 4: Selection efficiency dependence on the cut-off energy.

### 3.2.3 Effects from energy resolution and bias of energy scale

Event selection for the luminosity measurement is based on the criterion that the total deposited energy in the fiducial volume of the luminosity calorimeter is more than 80% of the center of mass energy. A possible uncertainty of the Bhabha selection efficiency due to the bias of measured energy, or the uncertainty of the stochastic term $\alpha_{res}$ in (2.1), will result in a corresponding uncertainty of luminosity measurement.

To keep the contribution to the luminosity uncertainty below $10^{-4}$, the absolute uncertainty of the measured deposited energy in the luminosity calorimeter (ΔE) would have to be 39 MeV. This results from the linear fit (Figure 4) at the energy cut-off, with the slope of -1.77·$10^{-3}$. Detector energy resolution must be understood to $\Delta\alpha_{res}$=2.5% to contribute to the relative uncertainty in luminosity of $10^{-4}$, as shown also in [10]. In the analysis above, photons radiated from the final state are excluded from the simulation. If radiative photons, emitted by Bhabha particle within the cone of one Moliere radius, are taken into account, $\Delta E$ and $\Delta \alpha_{res}$ are relaxed to 67 MeV and 4.3%, respectively [11].

### *3.2.4 Effects from polar angle resolution and bias in polar angle reconstruction*

The existence of the bias in polar angle reconstruction may cause a shift in the luminosity measurement, since events may be pushed in or out of the detector fiducial volume. With presently simulated bias of the polar angle reconstruction, a relative uncertainty of the luminosity measurement of $1.6 \cdot 10^{-4}$ is expected as the upper bound. In practice, it is possible to measure the bias of the polar angle reconstruction in a test beam. Only the uncertainty of such a measurement, will than contribute to the luminosity uncertainty. A contribution of similar size to the relative uncertainty of the luminosity measurement can be expected from the resolution of the polar angle reconstruction [11].

| Source of uncertainty | $\Delta L/L$ |
|---|---|
| Bhabha cross-section $\sigma_B$ | $5.4\ 10^{-4}$ |
| Polar angle resolution $\sigma_\theta$ | $1.6\ 10^{-4}$ |
| Bias of polar angle $\Delta\theta$ | $1.6\ 10^{-4}$ |
| Energy resolution $\alpha_{res}$ | $1.0\ 10^{-4}$ |
| Energy scale | $1.0\ 10^{-4}$ |
| Physics background B/S | $2.3\ 10^{-3}$ |
| BHSE | $1.5\ 10^{-3}$ |
| Beam polarization | $1.9 \cdot 10^{-4}$ |
| **Σ** | **$3.0\ 10^{-3}$** |

Table 1: Summary of systematic errors in the luminosity measurement. Errors are assumed to be uncorrelated. Uncertainty of the theoretical cross-section for Bhabha scattering is taken to be as at LEP energies.

### *3.2.5 Polarization of beams*

If polarization of electron and positron beams is available at ILC as foreseen, this will suppress the Bhabha cross section in the acceptance range of the luminometer by up to a few per cent [8]. In the current design, the maximum values for electron and positron polarization are 0.8 and 0.6, respectively, with an uncertainty of 0.0025 [9], producing a relative reduction of the Bhabha cross section of $2.3 \cdot 10^{-2}$ with an uncertainty of $1.9 \cdot 10^{-4}$ which in turn translates into a relative uncertainty of the luminosity mesurement.

## 4 Conclusion

At the present level of understanding of the detector performance and physics effects in luminosity measurement it has been shown that it will be possible to measure integrated luminosity at ILC with the total systematic uncertainty of $3 \cdot 10^{-3}$. The largest uncertainty due to two-photon background can clearly be reduced by correcting for it and using its uncertainty from NLO corrections as a true source of uncertainty of luminosity measurement. As well, effects from the polar angle reconstruction taken at present at full sizes will be replaced by their uncertainties once they are known.